
%
%
\font\tfont=cmmib10
\newfam\vecfam

\textfont\vecfam=\tfont \scriptfont\vecfam=\seveni
\scriptscriptfont\vecfam=\fivei


\def\s{\ifmmode \widetilde \else \~\fi}

\def\spose#1{\hbox to 0pt{#1\hss}} 
\def\lta{\mathrel{\spose{\lower 3pt\hbox{$\mathchar"218$}}
     \raise 2.0pt\hbox{$\mathchar"13C$}}}
\def\gta{\mathrel{\spose{\lower 3pt\hbox{$\mathchar"218$}}
     \raise 2.0pt\hbox{$\mathchar"13E$}}}



\def\msun{{\,M_\odot}}

\def\cm{{\rm\,cm}}

\def\g{{\rm\,g}}


\newcount\notenumber
\notenumber=1
\def\note#1{\footnote{$^{\the\notenumber}$}{#1}\global\advance\notenumber by 1}

\newcount\eqnumber
\eqnumber=1
\def\new{{\rm\chaphead\the\eqnumber}\global\advance\eqnumber by 1}
\def\eqref#1{\advance\eqnumber by -#1 \chaphead\the\eqnumber
     \advance\eqnumber by #1 }
\def\last{\advance\eqnumber by -1 {\rm\chaphead\the\eqnumber}\advance
     \eqnumber by 1}

\def\eq#1#2{\advance\eqnumber by -#2
\if#1(equation~(\chaphead\the\eqnumber\else eq.~[\chaphead\the\eqnumber\fi
\advance\eqnumber by #2
}
\def\eqs#1#2{\advance\eqnumber by -#2
\if#1(equations~(\chaphead\the\eqnumber\else eqs.~[\chaphead\the\eqnumber\fi
\advance\eqnumber by #2
}

\def\Eq#1#2{\advance\eqnumber by -#2
\if#1(Equation~(\chaphead\the\eqnumber\else Eq.~[\chaphead\the\eqnumber\fi
\advance\eqnumber by #2
}
\def\Eqs#1#2{\advance\eqnumber by -#2
\if#1(Equations~(\chaphead\the\eqnumber\else Eqs.~[\chaphead\the\eqnumber\fi
\advance\eqnumber by #2
}

\def\eqnam#1#2{\immediate\write1{\xdef\ #2{(\chaphead\the\eqnumber}}\xdef#1{
(\chaphead\the\eqnumber}}

\newcount\fignumber
\fignumber=1
\def\nfig{\the\fignumber\ \global\advance\fignumber by 1}
\def\nfiga#1{\the\fignumber{#1}\global\advance\fignumber by 1}
\def\rfig#1{\advance\fignumber by -#1 \the\fignumber \advance\fignumber by #1}
\def\fignam#1#2{\immediate\write1{\xdef\
 #2{\the\fignumber}}\xdef#1{\the\fignumber}}


\def\etal{{\it et~al.}}

\def\dotsfill{\leaders\hbox to 1em{\hss.\hss}\hfill}

\font\csc=cmcsc10   
\font\write=cmff10  

\font\ovtrm=cmr10 scaled 1200
\font\ovti=cmmi10 scaled 1200
\font\ovtsy=cmsy10 scaled 1200
\font\ovtex=cmex10 scaled 1000
\font\ovtit=cmti10 scaled 1200
\font\ovtsl=cmsl10 scaled 1200
\font\ovtbf=cmbx10 scaled 1200
\font\ovsrm=cmr7 scaled 1200
\font\ovsi=cmmi7 scaled 1200
\font\ovssy=cmsy7 scaled 1200
\font\ovsbf=cmbx7 scaled 1200
\font\ovfrm=cmr5 scaled 1200
\font\ovfi=cmmi5 scaled 1200
\font\ovfsy=cmsy5 scaled 1200
\font\ovfbf=cmbx5 scaled 1200


\def\foolit{\ifnum\pageno > 1 \number\pageno\fi}


\def\halfspace{\baselineskip=12pt plus .1pt}

\def\overhead{\def\rm{\fam0\ovtrm }
  \textfont0=\ovtrm   \scriptfont0=\ovsrm \scriptscriptfont0=\ovfrm
  \textfont1=\ovti   \scriptfont1=\ovsi \scriptscriptfont1=\ovfi
  \textfont2=\ovtsy   \scriptfont2=\ovssy \scriptscriptfont2=\ovfsy
  \textfont3=\ovtex   \scriptfont3=\ovtex \scriptscriptfont3=\ovtex
  \textfont\itfam=\ovtit  \def\it{\fam\itfam\ovtit }
  \textfont\slfam=\ovtsl  \def\sl{\fam\slfam\ovtsl }
  \textfont\bffam=\ovtbf \scriptfont\bffam=\ovsbf
\scriptscriptfont\bffam=\ovfbf
  \def\bf{\fam\bffam\ovtbf }
  \normalbaselineskip=36pt plus .1pt
  \setbox\strutbox=\hbox{\vrule height17.pt depth7.pt width 0pt}
  \normalbaselineskip\rm}

\def\papersize{\magnification=1200}  

\def\Pacz{Paczy\'nski\ }
\def\blankline{\par\vskip \baselineskip}



\parindent=40pt
\settabs 7 \columns
\tolerance=1600
\def\chaphead{}

\headline={\hss\tenrm\foolit\hss}    
\footline={\hfill}



\def\today{\csc\ifcase\month\or
  January\or February\or March\or April\or May\or June\or
  July\or August\or September\or October\or November\or December\fi
  \space\number\day, \number\year}

\def\ref#1{$^{#1}$}
\def\title#1\endtitle{\par\vfil\eject
     \par\vbox to 1.2in {}{\bf #1}\par\vskip 1.2in\nobreak}
\def\author#1\endauthor{\par{\csc #1}\par}
\def\institution#1\endinstitution{\par{\it #1}\par\blankline}

\def\section#1\endsection{\par\vfil\eject{\bf #1}\par\vskip 12pt\nobreak\rm}
\def\subsection#1\endsubsection{\vskip 14pt plus 50pt {\rm #1}\par
     \nobreak\blankline\nobreak\rm}

\parindent 2.5em
\parskip 1ex
\raggedbottom
\def\frac#1#2{{#1\over#2}}
\def\Mesz{M\'esz\'aros}
\def\newpage{ \par \vfill \supereject }

\def\ctl{\centerline}

\def\lbr{ \hfill\break }

\def\ref{\par \noindent \hangindent=2pc \hangafter=1 }
\def\etal{{\it et~al.}}
\def\mathnew{\mathsurround=0pt}
\def\simov#1#2{\lower .5pt\vbox{\baselineskip0pt \lineskip-.5pt
	\ialign{$\mathnew#1\hfil##\hfil$\crcr#2\crcr\sim\crcr}}}
\def\simg{\mathrel{\mathpalette\simov >}}
\def\siml{\mathrel{\mathpalette\simov <}}
\def\lambdabar{\mathrel{\lower 1pt\hbox{$\mathchar'26$}\mkern-9mu
        \hbox{$\lambda$}}}

\def\varep{\varepsilon}

\def\sla{\slash \kern -.75em}
\def\ssk{\vskip 1ex\noindent}
\def\msk{\vskip 2ex\noindent}
\def\bsk{\vskip 3ex\noindent}
\def\msun{M_{\odot}}

%
%
\def\cm{~\rm{cm}}

\def\s{~\rm{s}}

\def\g{~\rm{g}}

\def\cmsq{~ {\rm cm}^2 }

\def\cmcui{~ {\rm cm}^{-3} }

\def\eV{~\rm{eV}}
\def\keV{~\rm{keV}}
\def\MeV{~\rm{MeV}}
\def\GeV{~\rm{GeV}}

\def\erg{~\rm{ergs}}

\def\G{~\rm{G}}



\papersize
\halfspace
%
\ctl{\bf GASDYNAMICS OF RELATIVISTICALLY EXPANDING}
\ctl{\bf GAMMA-RAY BURST SOURCES:}
\ctl{\bf KINEMATICS, ENERGETICS, MAGNETIC FIELDS AND EFFICIENCY}
\bsk
\ctl{ P. \Mesz$^{1}$, P. Laguna$^{1}$ and M.J. Rees$^{2}$}
\bsk
\ctl{$^1$~ Pennsylvania State University, 525 Davey Lab, University Park, PA
16803}
\ctl{$^2$~ Institute of Astronomy, Madingley Road, Cambridge CB3 OHA, England}
\bsk
\ctl{ Submitted to Ap.J., Jan. 4, 1993}
\msk
{\bf Abstract}: We calculate both analytically and numerically the evolution of
 a
highly relativistic fireball through the stages of free expansion and
coasting, and determine the dependence of the thermodynamic and radiation
variables in the comoving and laboratory frames. Magnetic fields may have been
important in the original impulsive event. We discuss their effect on the
fireball dynamics,  and consider also their effects on the radiation
emitted when the fireball runs into an external medium
and is decelerated. The inverse synchro-Compton mechanism can then yield
high radiative efficiency in the reverse shock, producing a burst of
non-thermal radiation mainly in the MeV to GeV range whose total energy
and duration agree  with those of typical cosmic gamma ray bursts.
\bsk
%
\ctl{\bf 1.~ Introduction}
\msk

Gamma-ray burst sources (GRB) have long been suspected to originate from
the sudden release of energy in small regions of space, where the initial
energy density and characteristic photon energy is so large that an
opaque $e^\pm$ fireball forms (e.g. Cavallo and Rees, 1978, \Pacz, 1986,
Goodman, 1986, Shemi and Piran, 1990). If these events were located at
distances comparable to (or larger than) typical galactic scales, the $e^\pm$
fireballs would necessarily remain optically thick out to radii of at least
$\sim 10^9 \cm$ (e.g. Zdiarzki, 1982, Imamura and Epstein, 1987), depending on
the amount of normal electrons and baryons mixed in with the pairs. If the
fireball originated in a region smaller than this, the radiation pressure on
the optically thick fireball would cause it to expand; the evolution
of this fireball was initially thought to lead, when it became optically thin,
to the observed gamma-ray bursts. This model has recently received increased
attention because the spatial distribution of GRBs revealed by the BATSE
experiment
on the Compton Observatory  strongly suggested an extended galactic halo or a
cosmological origin (Meegan, \etal, 1991, Fishman, 1992, \Pacz, 1992).

In its simplest version, the fireball model failed to account for the duration
and time-structure of the observed bursts: it predicts a very
short burst, emitted when the expanding fireball becomes optically thin
(Goodman, 1986), and a quasi-thermal, soft $\gamma$-ray spectrum (Goodman,
1986,
\Pacz, 1986). Another problem, emphasized by \Pacz, 1990, was the possibility
that
the $\gamma$-rays could easily be degraded to even lower energies by adiabatic
cooling due to scattering on electrons associated with baryons polluting the
pair flow. In fact, in the latter case, most of the fireball energy gets
converted into bulk kinetic energy of the polluting baryons, while the photon
burst at thinning has much less energy than initially produced.
These problems are removed when one considers
anisotropic scenarios where high Lorentz factor fireballs can escape the
inevitable baryon-polluted slow wind that must accompany the initial energy
deposition (e.g. \Mesz~and Rees, 1992a, 1992b). In such cases, the main part
of the observed gamma-ray radiation occurs when the bulk kinetic energy of the
baryons (carrying essentially the full energy of the initial fireball) is
re-randomized and radiated away in the blast wave being pushed ahead of the
relativistically expanding baryons, and in the reverse shock that propagates
inwards into the baryonic fireball gas as the latter is decelerated by the
external medium (Rees and \Mesz, 1992, \Mesz~ and Rees, 1993). This is a
very generic mechanism, which operates in almost any scenario for the original
energy deposition, as long as it makes a relatively clean, high entropy
initial fireball. Moreover, the process depends on the external environment,
thereby allowing the possibility that even a standardised type of fireball
could
create   bursts with a variety of complex time-profiles.

The initial development of the fireball determines the final properties of the
blast waves, and thus of the bursts. In particular the final bulk Lorentz
factor, together with the external density, determines the blast wave burst
duration, while the dynamics of the fireball as a function of the baryon
loading determines the relative amount of energy in bulk kinetic form (which
gets radiated in the blast wave and reverse shock) and in pair and radiation
form (which escapes when the fireball thins, usually before the blast wave
burst). Since the relativistic dynamics of the fireball expansion determines
the entire energetics as well as the temporal characteristics of the burst,
a detailed calculation is important.

In the present paper we discuss the dynamics of the relativistic fireball
expansion from the acceleration stage through the coasting phase, calculating
both numerically and analytically the thermodynamic and radiation variables of
the flow through the transition to optical thinness and the saturation of the
bulk velocity. We discuss the kinematics both in the comoving and laboratory
frames, and indicate the scaling of the behavior with the various parameters
of the problem, extending the treatment to the anisotropic (jet) case.
We also consider the role of magnetic fields in the dynamics of the evolution,
including the case of magnetically dominated fireballs, and investigate the
effect of the magnetic field on the radiative efficiency of the reverse
shock that arises when the coasting fireball is decelerated in an external
medium.
\bsk
\ctl{\bf 2.~~ Numerical Treatment of a Highly Relativistic Expanding Gas}
\msk
{\it 2.1.~~ Computational Approach }
\ssk

Because of the presence of a natural length in the problem
(the initial value of the radius $r_o$) the gas dynamic equations do not
allow an exact similarity solution valid at all radii. For this reason, a
numerical solution is the only way to follow exactly the development of the
gas, although approximate analytic solutions are possible in the initial
acceleration stage and in the later coasting phase (see \S 3).
Previous calculations of the free expansion of a relativistic gas in
spherical symmetry (e.g. Vitello and Salvati, 1976, using a characteristics
method) have followed the evolution over one and a half decades in the
expansion
factor. However, expansion over such a limited range is usually not sufficient
for the bulk velocity to reach its ultimate saturation value, for the cases of
large values of the initial radiation to rest mass energy ratio $\eta$
considered
in many problems, such as the GRB fireball problem,
$$
\eta ={E_o \over M_o c^2} \simeq 10^3 E_{51} (M_o/ 0.6\times 10^{-6}\msun
)^{-1}~.
\eqno(2.1)
$$
Here $E_o \sim 10^{51}E_{51}\erg$ is the order of magnitude of the {\it
initial}
photon energy expected from the liberation of about a solar mass equivalent of
gravitational energy ($ G {\msun}^2/R_N \sim \epsilon\msun c^2 \sim 10^{54}
\erg$, where $\epsilon \siml 1,~R_n\sim 10^6\cm$, most of which goes into
neutrinos and gravitational waves), while $M_o$ is the total mass of
polluting baryons that get mixed in the photon-pair fireball of energy $E_o$.

The numerical treatment of the free expansion of a highly relativistic gas
with spherical symmetry calls for the use of Lagrangian hydrodynamics because
of the vast range of physical scales present in the problem, of the order
of ten or more decades.
We have developed a Lagrangian code for relativistic fluid flows
that uses a second order Runge-Kutta integrator (Benz 1984)
with adaptive time step. This method has the advantage that
a Courant (Courant, Friedrichs \& Lewy 1928) condition does
not limit the size of the time steps, an essential ingredient to
achieve short and large time scales.
Relativistic artificial viscous
stress was implemented in the code and calibrated
using relativistic shock tube problems.
This was used to verify the accuracy of the same code when artificial
viscosity was switched off, which turns to be quite good (and significantly
faster) in the cases of free expansion treated here. The reason why this
is an excellent approximation is that shocks do not normally occur in
such free expansion problems.  All the runs
described below are therefore for zero artificial viscosity.
Typically, simulations consisted of 200 Lagrangian
grid points (mass shells). Runs with different number of
mass shells were done to check the convergence of our results.
We monitored the ability of the code to reproduce
the analytic solutions of the plane symmetric rarefaction wave traveling
inwards (Vitello \& Salvati 1976).
We found that for 200 grid points the
maximum absolute error ($\le 3 \%$) as well as the
maximum cumulative error ($\le 2 \%$) appeared in the velocity.
\bsk
{\it 2.2~~ Numerical Results}
\ssk

The initial gas configuration adopted here, for simplicity, is that the
gas at $t= 0$ is uniformly distributed within $r \leq r_o$, having a
total rest mass $M_o$, and that at $t=0$ an amount of radiation energy
$E_o \gg M_o c^2$ is deposited uniformly throughout the spherical gas.
The gas is optically thick to its own electrons, and the resulting
photon-pair-electron-baryon fluid is essentially an isentropic fluid which
can be modeled as a gas of adiabatic index $\gamma_a= 4/3$, as long as the
photons are coupled to the matter by radiation drag. The same applies to the
case where the energy is mainly in the form of magnetic fields (see \S4).
After decoupling, the adiabatic index becomes $\gamma_a=5/3$. We have
carried out calculations for values of $\eta$ ranging from close to unity
up to $10^{10}$. As a specific example, we discuss the case of $\eta=10^4$
below, which is characteristic of the relativistic behavior.

In the course of the expansion most of the mass, as seen by a laboratory
observer, is concentrated in a thin shell near the leading edge of the
expanding gas (c.f. also Vitello and Salvati, 1976). This is seen in Fig. 1,
where the distribution of the lab density is plotted as a function of the
(normalized) lab radius for $\eta=10^4$. The curves are plotted at
the times (from top to bottom) when the lab frame expansion factor
$r(\rho_{mx})/r_o $ (which is proportional to the lab frame time) is equal to
$0.2,~0.4,~0.6,~0.8,~1.0$ times $\eta$.

As the gas expands, the mass shells acquire a bulk velocity which initially
increases in time and eventually saturates. The distribution of this
bulk Lorentz factor over the various mass shells is shown in Fig. 2, for
a fireball of $\eta=10^4$ at the lab times (increasing from bottom to top) when
$r(\rho_{max})/r_o = \eta^{1/2}, \eta, \eta^{3/2}, \eta^2, \eta^{5/2}, \eta^3$.
One sees that the leading edge reaches a value of
$\Gamma \sim \eta$, and most of the mass reaches one third of that value,
at the laboratory time for which the expansion factor $r(\rho_{mx})/r_o \sim
\eta$. After that the average bulk Lorentz factor of most of the matter
saturates to the value $\Gamma \sim \eta$, while the leading edge remains
within a factor 2 of that value, and the trailing edge of the inner 10\% of
the matter tapers off fairly steeply to zero.

The average bulk Lorentz factor (or what is nearly the same,
$\Gamma(\rho_{max})$ where the density is maximal)
grows initially linearly with radius, and saturates to a value
$\Gamma\sim \eta$ after a (lab) expansion factor equal to $\eta$. This is
seen in Fig. 3, where $\Gamma(\rho_{max})$ is plotted against the expansion
factor $r(\rho_{max})/r_o$.

Within the thin shell containing most of the mass, called the mass envelope
shell, the bulk of the mass (say the inner 80\% away from the outer and
inner edges) is distributed quite uniformly, even after the
average bulk Lorentz factor has saturated (for $r(\rho_{max})/r_o \simg
\eta$).  Only later, particularly after $r(\rho_{max})/r_o \simg \eta^2$,
a slight asymmetry favoring the leading edge of the envelope shell starts to
become apparent, but even then the matter within the shell can be considered
uniform to a good approximation. This is shown in Fig. 4a, where the lab frame
density $\rho_L$ is plotted against the Lagrangian mass coordinate $M/M_t$
(where $M_t\equiv M_o$ is the total rest mass, which is mostly within the
shell).
In figure 4b we show the same plot for the comoving mass density $\rho_c$.
The latter is even more uniform than the laboratory frame density. Thus,
a comoving observer would find that the world around it is essentially
isotropic and homogeneous, at least far from the edges, but with a density
that drops in time, as seen from the decreasing value of the average density.
The various curves, from top to bottom, are for (increasing) times when the
expansion factor $r(\rho_{mx})/r_o $ is equal to $\eta^{1/2}, \eta,
\eta^{3/2}, \eta^2, \eta^{5/2}, \eta^3$.

The width of the mass envelope shell in the lab frame is initially nearly
constant an equal to the initial radius, $\Delta r/r_o \sim 1$
(c.f. also Vitello and Salvati, 1976, Goodman, 1986). This remains so,
however, only until an expansion factor $r(\rho_{max})/r_o \sim \eta^2$
is reached. After that, the lab width starts to grow linearly (see Fig. 5a).
The comoving frame width, for its part, grows initially linearly with
the expansion factor until $\eta$, remains approximately constant between
$\eta$ and $\eta^2$, and then resumes a linear growth with radius (Fig. 5b).
The physical reason for this behavior is discussed in \S 3.

We have also calculated numerically the lab radius at which an expanding
fireball
becomes Thomson optically thin against its own electrons, as a function of the
initial radiation energy to rest mass ratio $\eta$. This condition is
computed in the comoving frame, being the same in the lab frame, as it
should, since the optical depth of a constant amount of mass is an
invariant between reference frames. This (lab) thinning radius $r_t$ is
found to vary as $\eta^{-1/2}$, in excellent agreement with the analytic
estimate
(3.13). Another quantity of interest is the value of
average bulk Lorentz factor at which fireballs of varying $\eta$ become
optically thin, as a function of $\eta$. The numerically computed value of
$\Gamma_t$ grows linearly with $\eta$ for low values of $\eta$ where optical
thinness is achieved before the bulk velocity $\Gamma$ saturates to $\eta$,
and $\Gamma_t$ decreases as $\eta^{-1/2}$ for larger values of $\eta$ for
which thinness is achieved after the bulk Lorentz factor has saturated (in
agreement with Shemi and Piran, 1990). The linear decrease $\propto
\eta^{-1/2}$
changes to $\eta^0$ for very large values of $\eta$ (or initial radiation
energies sufficiently larger than $M_o c^2$) at which the opacity of the
fireball
at optical thinness is still dominated by the pairs (instead of by the
polluting baryonic electrons). This occurs for $\eta \simg 0.6\times 10^{10}$,
above which $r_t \sim$ constant, $\Gamma_t\sim 2.4\times 10^3\sim$ constant
(see \S 3).
\bsk
\ctl{\bf 3.~~ Analytical Treatment of the Fireball Dynamics}
\msk
{\it 3.1~~Accelerated and Saturated Expansion}
\ssk

A fireball expanding into a low density external environment may, in its
initial stages, be considered to be expanding in a vacuum. Thus, as long
as the inertia of the accumulated external matter can be neglected, shocks
will not play an important role in the energetics. Pressure gradients may
exist in the expanding fireball gas, particularly if the initial density and
pressure distribution is inhomogeneous, but a free (unimpeded) expansion
will tend to stretch these out. One may therefore, as a first approximation,
neglect internal pressure gradients. Thus, even though we are dealing with a
fluid, the dynamics of the expansion would be expected initially to resemble
that of a collisionless gas of relativistic particles. The particles
initially have an isotropic velocity distribution within a radius $r_o$, with
an initial thermal Lorentz factor $\eta=E_o/M_o c^2$. But as they expand, the
velocity distribution viewed in the lab frame will become increasingly
anisotropic. For the ballistic expansion in the lab frame, when the particles
have reached a radius $r$ their velocity vectors will be confined within
an angle $(r/r_o)^{-1}$ of the radial direction. A transformation to a frame
of reference which is outwardly moving with a Lorentz factor
$$
\Gamma \sim \theta^{2/3}(r/r_o)~, \eqno(3.1)
$$
makes the particle velocity distribution appear again isotropic.
This frame is, therefore, the comoving frame of the expanding particles
in the initial acceleration phase, which moves outward with a velocity
increasing according to $\Gamma\propto r$. (We have introduced in (3.1)
an angular factor $\theta$ which approximately accounts for jet geometry;
for spherical symmetry $\theta \sim 1$, see below).

The expansion occurs at the expense of the comoving
frame thermal energy $E$, which from conservation of energy must therefore
drop as $E\propto r^{-1}$. This agrees with the adiabatic expansion law
$E\propto V^{-1/3}\propto \rho^{1/3}$, where $V,~\rho$ are comoving volume
and comoving baryon density, provided that the comoving volume
$V\propto r^3$. The latter follows from the
fact that the laboratory frame radial extent of the particles in
free expansion is expected to be (initially, at least) $\Delta r\sim r_o$,
while the comoving radial width $\Delta R$ is related to the corresponding
lab width through $\Delta R = \Delta r \Gamma$ (we denote comoving radii
with capital $R$ and laboratory radii with lower case $r$). Since
the dimensions transverse to the motion will be the same in the lab
and comoving frame, the comoving volume in this (acceleration) phase is
$V\propto r^2 \Delta R \propto r^2 r_o\Gamma \propto r^3$.

The accelerating behavior $\Gamma \propto (r/r_o)$, however, can only go
as long as the internal energy of the expanding gas is
relativistic. After the comoving energy density drops below the baryon rest
mass density $\rho c^2$, the bulk Lorentz factor $\Gamma$ must saturate to
the maximum value it can acquire, which is approximately the initial thermal
Lorentz factor $\eta$ in the case of significant baryon loading, or more
generally
$$
\Gamma_s\sim \hbox{ min}~[ \eta ~,~ \Gamma_m ]~, \eqno(3.2)
$$
where the second value is appropriate for $\eta >\Gamma_m$ given in eq.(3.14),
as shown in \S 3.4.

The initial acceleration and the saturation behavior can also be obtained
from a phenomenological expression for the average bulk Lorentz factor
(e.g., Shemi and Piran, 1990)
$$
\Gamma \sim {E_o + M_o c^2 \over E + M_o c^2}={{\eta +1}\over {\eta(E/E_o)+1}}
$$
making the analogy with a section of the expanding universe, i.e. assuming
spherical symmetry, homogeneity and isotropy. Here then $E$ is the fluid
thermal energy density (initially mainly in photons and leptons, with an
admixture of $M_o$ baryons). As long as it is optically thick and $\eta \simg
1$, the pressure is radiation-dominated throughout the expansion, and
using the adiabatic behavior of a relativistic fluid one has
$$
E/E_o=T/T_o= (\rho/\rho_o)^{1/3}= (V/V_o)^{-1/3}~, \eqno(3.4)
$$
where $E,~T,~V,~\rho$ are comoving thermal energy, temperature, volume and
baryon density. In the linear expansion phase, we can take (from the discussion
below equation (3.1)) $V^{1/3}\propto r^{-1}$, where $r$ is lab radius, so
 From (3.3) and (3.4) one gets for $E_o/E \ll \eta$ the accelerating behavior
$\Gamma \sim E_o/E \sim r/r_o$, while for $E_o/E \gg \eta$ one gets the
saturated $\Gamma \to \eta$, c.f. eq.(3.2). The saturation occurs (in
spherical symmetry) at $r_s/r_o \sim \eta$. However, for $\eta >\Gamma_m$ given
by eq.(3.14), $\eta$ must be replaced by $\Gamma_m$, see \S 3.4.
\bsk
{\it 3.2~~ Lab and Comoving Geometry}
\ssk

As the particles move outward with velocity vectors which are increasingly
radial, they form a radially expanding shell whose radius is initially
$\Delta r\sim r_o$. The radial velocity spread $(v-c)/c\sim \Gamma^{-2}$,
and a noticeable departure from the approximately constant width $r_o$ starts
to become appreciable only after $r\simg r_b$ where $\Delta r \sim r
\Delta v/c \sim r_b \eta^{-2} \simg r_o$, or $r_b /r_o \simg\eta^2$.
The laboratory frame width is therefore
$$
\Delta r \sim ~\hbox{max}~ [r_o~,~ r/\Gamma^2 ]~\sim~ \cases{
r_o~,& ~~~~for $r \siml r_b$;\cr
r/\Gamma^2~,& ~~~~for $r \simg r_b$ ~.\cr}   \eqno(3.5)
$$
The width in the comoving frame is $\Delta R = \Delta r \Gamma$, which from
eqs. (3.1),(3.2) and (3.5) is
$$
\Delta R \sim \cases{
\theta^{2/3} r~& ~~~~for $r\siml r_s$; \cr
r_o\Gamma_s~& ~~~~for $r_{s} \siml r \siml r_b$; \cr
r/\Gamma_s~& ~~~~for $r \simg r_b$~,\cr}      \eqno(3.6)
$$
where the two characteristic radii for saturation and for the start of the
shell
linear expansion in the lab frame are
$$
r_{s}/r_o \sim \theta^{-2/3}\Gamma_s~~~~,~~~~r_b/r_o \sim \Gamma_s^2~.
\eqno(3.7)
$$
The comoving volume in the three different stages of free expansion is
$$
V=4\pi \theta^2 r^2\Delta R \sim \cases{
4\pi\theta^2 r^3 ~&~~~~for $r <r_s$;\cr
4\pi\theta^2 \Gamma_s r_o r^2 ~&~~~~for $r_s <r< r_b$;\cr
4\pi\theta^2 \Gamma_s^{-1} r^3 ~&~~~~for $r > r_b$;\cr} \eqno(3.8)
$$
and the comoving radiation energy and temperature vary according to
$$
({E\over E_o})=({T\over T_o})=\cases{
\theta^{-2/3} (r_o/r) ~&~~~~for $r <r_s$;\cr
\theta^{-2/3}(r_o/r_s)(r_s/r)^{2/3}=\theta^{-4/9}\Gamma_s^{-1/3}(r_o/r)^{2/3}
                                              ~&~for $r_s <r< r_b$;\cr
\theta^{-2/3}(r_o/r_s)(r_s/r_b)^{2/3}(r_b/r)=\theta^{-4/9}\Gamma_s^{1/3}(r_o/r)
                                             ~&~for $r > r_b$;\cr} \eqno(3.9)
$$
This is therefore the complete $r$ dependence of the adiabatic law (3.4).
The factor $\theta$ can be taken to include both an angle dependence and
a possible dependence on the statistical weight factor $g$ in the equilibrium
energy density $gaT^4$. That is,
$\theta^{-2/3}\equiv (g_o/g)^{1/3}\theta'^{-2/3}$, where
$(g_o/g)^{1/3}= (11/4)^{1/3}$ accounts for the change in the statistical
factor for the equilibrium energy density when the pairs annihilate as the
temperature drops below $kT\siml m_ec^2$, and $\theta'^{-2/3}$ accounts for
the possibility of expansion along a restricted range of solid angles.
If we have radial expansion along two jets of solid angle $\theta_o$, and
all of the energy is channeled into this solid angle, then $\theta' \sim
\theta_o 2^{-1/2}$ (or $\sim \theta_o /2$ if there is one jet only). If the
gas uses some its internal energy to perform work against a medium which
prevents escape in the directions outside of the jet angles $\theta_o$, then
the effective $\theta' < \theta_o 2^{-1/2}$. In the discussion below (and
in the numerical calculations of \S 2) we have ignored
the small shift of the curves associated with the change of the statistical
factor, i.e. we simplify to $(g_o/g)\sim 1,~\theta\sim \theta'$.
(Note that this angular factor is only approximately correct, since it does not
account for any possible lateral expansion or transverse radiation loss of
the jet; it is strictly valid for one-dimensional radial variations within
the angle $\theta'$).
For a jet, the asymptotic accelerated behavior has the same radial dependence,
but displaced to higher radii by $\theta^{-2/3}$, as is the saturation radius,
$$
\Gamma\sim \theta^{2/3}(r/r_o)~~~,~~~ r_{s}/r_o \sim \theta^{-2/3} \eta
{}~~,~~~ \Gamma_{s}\sim \eta~,.  \eqno(3.10)
$$
(which is valid only for $\eta <\Gamma_m$; otherwise
$\eta$ must be replaced by $\Gamma_m$ given by eq.(3.14) see \S 3.2). The
effect
of the approximate angular factor, e.g. in the behavior of the growth of
the bulk Lorentz factor (eq.(3.1)), is that,
for the same initial energies the jet achieves the same velocities
as a corresponding spherical flow, but at larger radii. This is because at the
same radii the jet, being confined to narrower angles, has a larger internal
entropy than the spherical flow. The saturation Lorentz factor is however
the same in both cases.
\msk
{\it 3.3~~ Optical Depth and Saturation Lorentz Factor}
\msk

Depending on the value of $\eta=E_o/M_oc^2$, the scattering depth of
the fireball at optical thiness is dominated by the ``polluting" baryonic
electrons (if $\eta<\eta_p$, where the latter is given by eq.(3.19)), or, if
the
fireball has very low baryon pollution, the scattering opacity previous
to reaching optical thinness is dominated by the $e^\pm$ pairs (Goodman, 1986,
\Pacz, 1986, Shemi and Piran, 1990), for $\eta >\eta_p$.
\msk
a) Electron-dominated scattering: In the case $\eta <\eta_p$, when baryonic
electrons dominate at thiness, the optical depth is
$$
\tau={M_o/m_p \sigma \Delta r \over 4\pi r^2 \theta^2 r^2 \Delta r}=
\bigl({E_o \kappa \over 4\pi r_o^2 c^2}\bigr)\bigl({r_o\over r}\bigr)
                                                {1\over \theta^2 \eta} =
\theta^{-4/3}{\Gamma_m^3\over \eta}\bigl({ r_o \over r}\bigr)^2~.
                                                               \eqno(3.11)
$$
Here $\sigma$ is scattering cross section, $\kappa=\sigma/m_p\sim 0.4 \cmsq
\g^{-1}$,
and $\Gamma_m$ is defined in eq.(3.14). The initial optical depth is $\tau_o=
\Gamma_m^3\theta^{-4/3} \eta^{-1}= \Sigma_o\kappa=
\eta^{-1}\Sigma_{r,0}\kappa$, where
$$
\Sigma_o\equiv M_o/4\pi r_o^2 \theta^2~~~,~~~
\Sigma_{r,o}\equiv E_oc^{-2}/4\pi r_o^2 \theta^2 ~,\eqno(3.12)
$$
are the initial baryon mass surface density and the initial radiation
equivalent
mass surface density. In the course of the expansion, the gas becomes optically
thin at a radius $r_t$ defined by
$$
{r_t \over r_o} = \theta^{-3/2} \Gamma_m^{3/2} \eta^{-1/2}
       = (\Sigma_o \kappa )^{1/2} =\tau_o^{1/2}
       = 1.9\times 10^8 E_{51} r_6^{-1}\theta^{-1}\eta^{-1/2}~,\eqno(3.13a)
$$
or
$$
r_t=\bigl({E_o \kappa \over 4\pi \theta^2 c^2 \eta}\bigr)^{1/2}=
$$
where the initial energy $E_o$ and radius $r_o$ have been arbitrarily
normalized to
$10^{51}\erg$ and $10^6\cm$, and the second version (3.13b) is independent of
$r_o$.
The critical Lorentz factor $\Gamma_m$ and the critical $\eta_m$ are defined as
$$
\eqalignno{
\Gamma_m\equiv\eta_m &
  = \theta^{4/9}(\tau_o \eta)^{1/3}=\theta^{4/9}(\Sigma_o\kappa\eta)^{1/3}
                                 =\theta^{4/9}(\Sigma_{r,o}\kappa)^{1/3} & \cr
        & =\bigl({ E_o \kappa \over 4\pi r_o^2 c^2}\bigr)^{1/3}\theta^{-2/9}
   =~ 3.3\times 10^5 ~E_{51}^{1/3}r_6^{-2/3}\theta^{-2/9}~. & (3.14) \cr}
$$
This is the maximum possible bulk Lorentz factor achievable for a given initial
radiation energy $E_o$ deposited within a given initial radius $r_o$ (\S 3.4).
It is also, for a given $E_o$ and $r_o$, the ``critical" value of $\eta=\eta_m$
(reached at a critical loading mass $M_o=M_m=(E_o/\eta_m c^2)$) for which the
thinning radius (3.13) is equal to the saturation radius $(r_s/r_o) \sim
\theta^{-2/3}\eta$. For a given $E_o$, as one varies $M_o$ or $\eta$, the
critical
value $\eta =\eta_m=\Gamma_m$ is reached at a radius $r_m$ given by
$$
{r_m\over r_o}={r_s \over r_o}={r_t\over r_o} \sim \theta^{-2/3}\Gamma_m~.
\eqno(3.15)
$$
For $\eta <\Gamma_m$, optical thinness is reached after saturation, whereas
for $\eta >\Gamma_m$ optical thinness is reached before saturation, so the
bulk Lorentz factor at thinning $\Gamma_t=\theta^{2/3}(r_t/r_o)$ is equal to
$$
\Gamma_t ~= ~\bigl[ \eta~,~ {\Gamma_m^{3/2} \eta^{-1/2}} \bigr]~~~~
\hbox{ for}~ [ \eta <\Gamma_m~,~\Gamma_m < \eta <\Gamma_p ]~,\eqno(3.16)
$$
where $\Gamma_p$ is given by eq.(3.20).
\bsk
b) Pair-dominated Scattering:
For very large $\eta$, such that $\eta >\eta_p > \Gamma_m$, the $e^\pm$
scattering dominates, instead of baryonic electron scattering, and optical
thiness occurs when the pairs fall out of equilibrium, which occurs at a lab
radius $r_p$ given by
$$
{r_p \over r_o}= \theta^{-2/3}{T_o \over T_p}\sim 2.4\times 10^3 E_{51}^{1/4}
r_6^{-3/4}\theta^{-2/3}~,\eqno(3.17)
$$
where $T_p\sim 15\keV$ is the temperature where the thermal pair density
becomes negligible, and $T_o\sim 36.4 E_{51}^{1/4}r_6^{-3/4}\MeV$ is the
initial temperature. This occurs during the acceleration stage, so at
this point, where pairs become optically thin, $\Gamma$ has reached the value
$$
\Gamma_t\equiv\Gamma_p =E_o/E_p=T_o/T_p=\theta^{2/3}(r_p/r_o)
           \sim 2.4\times 10^3 E_{51}^{1/4}r_6^{-3/4}~,  \eqno(3.18)
$$
valid for $\eta>\eta_p$. The latter is the value of $\eta$ above which the
pair-dominated regime occurs,
$$
\eta_p =\Gamma_m^3 /\Gamma_p^2 \sim 0.63\times 10^{10}E_{51}^{1/2}r_6^{-1/2}
\theta^{-2/3}~,  \eqno(3.19)
$$
in terms of which one can write the Lorentz factor and the radius at which
pairs become optically thin as
$$
\Gamma_p ={\Gamma_m^{3/2} \over \eta_p^{1/2} }~~~~~,~~~~~ {r_p\over r_o}=
\theta^{-2/3}\Gamma_p =\theta^{-2/3}{\Gamma_m^{3/2}\over \eta_p^{1/2}}~.
                                                          \eqno(3.20)
$$
\msk
{\it 3.4~~ Photon Drag and Final Baryon Lorentz Factor}
\ssk

Photons are obviously coupled to the baryons when $\tau >1$, which ensures
a radiation-like equation of state with adiabatic index 4/3. However, even
after $\tau <1$, baryons may remain coupled to the photons if the density
of the latter is so large that the Compton drag time is shorter than the
expansion time in the comoving frame. The comoving Compton drag time is the
time
during which an electron sweeps up an amount of photons whose mass equivalent
equals one proton plus one electron's mass, $t_D\sim m_pc^2/\sigma_T c
u_\gamma$,
where $u_\gamma$ is the comoving photon energy density. This is longer than
the Compton cooling time by a factor $m_p/m_e$.
The ratio of the comoving baryon rest mass energy density $u_m$ and the
comoving photon energy density $u_\gamma$ is
$$
\varepsilon \equiv {u_\gamma \over u_m}= {E\over M_o c^2}=\eta {E\over E_o}~,
\eqno(3.21)
$$
which is equal to $\theta^{-2/3}\eta r_o/r$ if $r<r_s$.
Dividing the Compton drag time $t_D$ by the comoving expansion time
$\Delta R /c$ and multiplying and dividing by the comoving
baryon density $n$ (comoving baryon rest energy density $u_m$), the ratio of
the Compton drag time to the expansion time is
$$
\zeta \equiv {t_D\over t_E}={ u_m \over u_\gamma ~\tau}=
{1\over \varepsilon ~\tau}=
({E_o \over E}) {\theta^{4/3} \over \Gamma_m^3}({r\over r_o})^2 =
\theta^2 \Gamma_m^{-3} \bigl({r\over r_o}\bigr)^3~, \eqno(3.22)
$$
where $\tau$ is the optical depth (3.11), and the last equality is valid
for $r \siml r_s$, with different dependences for $r \simg r_s$. The
final decoupling of photons and baryons occurs at a radius $r_f$ where
$\zeta=1$, or at the thinning radius $r_t$, whichever is largest.

The final bulk Lorentz factor of the baryons is the value that it has at
decoupling from the photons, either when the fireball becomes optically thin
or when the Compton drag has become longer than the expansion time, $\zeta >1$,
whichever occurs last. For $\eta <\Gamma_m$, the fireball saturates to a
value $\Gamma =\eta$, and thinning occurs at a radius larger by
a factor $(r_t/r_s)=(\Gamma_m /\eta)^{3/2}>1$ where
$\zeta_t=(\Gamma_m/\eta)>1$,
so the final baryon Lorentz factor is $\Gamma_f \equiv \eta$ in this case.
On the other hand, for $\eta >\Gamma_m$, at optical thinness $\varep_t=
(\eta/\Gamma_m)^{3/2}>1$
and $\zeta_t=(\Gamma_m/\eta)^{3/2}<1$, so the Compton drag keeps photons and
baryons coupled beyond $r_t$. This means that, even though most photons do
not collide, most electrons do keep colliding with a small fraction of the
photons, and keep being accelerated (with their baryons). They finally decouple
when $\zeta_f=1$ at $(r_f/r_o)=\theta^{-2/3}\Gamma_m$, where $\Gamma_f \sim
\Gamma_m$, $\varepsilon_f= (\eta/\Gamma_m)>1$ and $\tau_f=(\Gamma_m/\eta)<1$.
The terminal baryon bulk Lorentz factor is therefore
$$
\Gamma_f = ~\hbox{min}~\bigl[ \eta~,~\Gamma_m \bigr]~~~,~~~
\hbox{for}~\bigl[ \eta \leq \Gamma_m~,~\eta\geq \Gamma_m \bigr]~.\eqno(3.23)
$$
This is equal to the Lorentz factor at thinning $\Gamma_t$ in the
case $\eta <\Gamma_m$, when $\Gamma_t=\Gamma_f=\eta$, but $\Gamma_t\neq
\Gamma_f$ for $\eta>\Gamma_m$.
\msk
{\it 3.5~~ Baryon Loading and Final Radiation to Kinetic Energy Ratio}
\ssk

One can distinguish four baryon-loading regimes, characterized by the
value of $\eta$: \lbr
H) High-load fireballs , for $1 < \eta <\Gamma_m$, or $0.6\times 10^{-3}
\msun \siml M_o \siml 1.8\times 10^{-9}\msun$. In this regime $\Gamma$ grows
linearly with $r/r_o$ until reaching $r_s/r_o=\theta^{-2/3}\eta$, where it
saturates to $\eta$, and becomes optically thin at $r_t/r_o =\Gamma_m
(\Gamma_m /\eta)^{1/2} >\Gamma_m > r_s/r_o$.
The observer-frame (Doppler blueshifted) energy of the radiation
escaping under the assumption of isotropy (even if in reality it is beamed)
and the kinetic energy of the baryons is
$$
E_{r,ob}=E_o \theta^{-4/9}\eta^{-1/3}({r_o \over r_t})^{2/3}=
E_o({\eta\over \Gamma_m})<E_o ~~,~~E_k=M_oc^2\eta\simeq E_o~.  \eqno(3.24)
$$
That is, the thinning radiation burst is a small fraction of $E_o$ (unless
$\eta \to \Gamma_m$), most of the energy having gone into the kinetic
energy of the baryons.
\lbr
M) Critical-load Fireballs, for $\eta =\Gamma_m$ or $M_o\sim 1.8\times
10^{-9}\msun$. In this case $\Gamma_s=\Gamma_m$ at $r_s=r_t=r_f$, and both
the observer-frame radiation energy and the baryon kinetic energy assume
their maximal value, $E_{r,ob} \sim E_o/2$, $E_k\sim E_o/2$.
\lbr
L) Low-load or underloaded fireballs, for $\Gamma_m <\eta <\eta_p$, or
$1.8\times
10^{-9}\msun \siml M_o \siml 0.95\times 10^{-13}\msun$. For these (rather low)
values of the baryon pollutant mass, $\Gamma$ grows linearly with $r/r_o$ until
becoming optically thin at $r_t/r_o <r_s/r_o \sim r_m/r_o$, where it has the
value
$\Gamma_t \sim \theta^{2/3} r_t/r_o= \Gamma_m (\Gamma_m/\eta)^{1/2} <\eta$.
Most of the radiation
energy escapes from here without further scattering, while the baryons
continue coupled by Compton drag to a small fraction of the photons until
the radius $r_f$ where $\Gamma_f=\Gamma_m$. The bulk of the
observer-frame radiation energy observed (assumed over $4\pi$) and
the final kinetic energy of the protons are
$$
E_{r,ob}\sim E_o \theta^{-2/3}({r_o\over r_t})\Gamma_t \sim E_o~~,~~
E_k=M_oc^2\Gamma_f \sim E_o ({\Gamma_m\over\eta})< E_o~.  \eqno(3.25)
$$
P) Pair fireballs, for $\eta >\eta_p$ or $M_o \siml 0.95\times 10^{-13}\msun$.
In this extremely underloaded regime, the pairs dominate the Thomson opacity
and the value of $\Gamma$ grows linearly with $r/r_o$ but it always becomes
optically thin at the same value of the radius where $\Gamma\sim \Gamma_p$,
$r_t/r_o =r_p/r_o \sim \theta^{-2/3}\Gamma_p$, given by eqs.(3.18)-(3.20),
independent of $\eta$ as long as the latter is greater than $\eta_p$ .
The observer-frame radiation energy at thiness under the assumption
of isotropy and the final kinetic energy of the baryons is again given by
eq.(3.25). \lbr
The significance of the ratio of observed radiation energy to kinetic energy,
in our ``standard" model (\Mesz~ and Rees, 1993), is that this represents the
ratio of the energy in a short precursor burst to that in the main burst,
the latter coming from the recovery of the kinetic energy upon interaction with
the
external medium (\S 4).
\bsk
\ctl{\bf 4.~~ Magnetic Fields, Efficiency and Photon Energies}
\bsk
{\it 4.1~~ Magnetic Fireballs}
\ssk

In almost any model of the initial energy release, the initial mass motions
are expected to be extremely violent ($v \sim c$), and this could magnify any
pre-existing magnetic fields, via compression, shearing, turbulent dynamo
mechanisms, Parker-type instabilities, etc. (e.g. Usov, 1992, Narayan, \Pacz~
and Piran, 1992, Thompson and Duncan, 1993). If the initial total ``disposable"
energy (i.e., that portion of the available gravitational energy that is not
lost
in the form of neutrinos) is $E_o=10^{51}E_{51} \erg$, this might be
distributed between radiation and magnetic components as
$$
E_{Bo}=\xi E_o~~~,~~~E_{ro}=(1-\xi)E_o~,\eqno(4.1)
$$
where $\xi\leq 1$. One may define separate magnetic and radiation $\eta$
parameters
$$
\eta_B=E_{Bo}/M_oc^2=\xi\eta~~~,~~~\eta_r=E_{ro}/M_oc^2=(1-\xi)\eta~\eqno(4.2)
$$
in terms of the usual total $\eta=E_o/M_oc^2$, with $\eta=\eta_B~+~\eta_r=
\xi\eta~+~(1-\xi)\eta$. Even if the magnetic fields have a large scale
structure,
the expansion is essentially isotropic in the comoving frame, and therefore the
magnetic energy density will evolve $\propto V^{-4/3}$, so the total magnetic
energy and radiation energy in the comoving frame vary in a similar manner,
$E_B/E_{Bo}=B^2/B_o^2=E_r/E_{ro}=(V/V_o)^{-1/3}$, where the volume factors are
given by eqs. (3.9). The bulk Lorentz factor, similarly to (3.3), is again
$$
\Gamma={ E_o+M_o \over E_r +E_B+M_o}=
{\eta+1 \over \eta_r(E_r/E_{ro})+ \eta_B(E_B/E_{Bo})+1}=
{\eta+1 \over \theta^{-2/3}(r_o/r)\eta +1}~,\eqno(4.3)
$$
so that, as in the case of pure radiation, $\Gamma \sim \theta^{2/3}(r/r_o)$
for $r\siml r_s$ while $\Gamma\sim \eta$ for $r\simg r_s$, with $r_s/r_o=
\theta^{-2/3}\Gamma_s$ and $\Gamma_s=\hbox{min}[\eta, \Gamma_m]$, where
$\Gamma_m$ is given by eqs(3.14). The behavior is therefore the
same as in \S 3, even if all the disposable energy $E_o$ goes into
magnetic fields ($\xi\to 1$), i.e. a purely magnetic fireball.
The maximal initial magnetic field will be
$$
B_o=(8\pi \xi E_o /V_o)^{1/2}=10^{17}\xi^{1/2} E_{51}^{1/2} r_6^{-3/2}\G~,
                                                                  \eqno(4.4)
$$
and the comoving field strength evolves according to
$$
({B\over B_o})=\cases{
\theta^{-4/3} (r_o/r)^2 ~&~~~~for $r <r_s$;\cr
\theta^{-8/9}\Gamma_s^{-2/3}(r_o/r)^{4/3} ~&~for $r_s <r< r_b$;\cr
\theta^{-8/9}\Gamma_s^{2/3}(r_o/r)^2 ~&~for $r > r_b$;\cr} \eqno(4.5)
$$
where $r_s,~r_b$ are given in (3.7). Even after most of the pairs annihilate,
the exponentially small fraction of frozen-in surviving pairs is enough to
provide the currents needed to support large-scale fields, so that the behavior
(4.5) is uninterrupted beyond the radius at which annihilation, optical
thinness,
photon drag decoupling, etc. occur.

We may also consider briefly the most extreme scenario of magnetic field
dominance, that where the initial field $B_o$ is in equipartition not with
the disposable energy $E_o\sim 10^{51} E_{51}\erg$ (which, as in
supernovae, is of order $10^{-3}$ of the total liberated binding energy from
the gravitational collapse of about a solar mass) but rather is in
equipartition with the binding energy itself, $E_b\simeq G \msun^2/R_N \sim
10^{54} E_{54} \erg$. This might occur in a rapidly-spinning object, where
rotational and shearing motions would involve a large fraction of the total
binding energy. In this case
$$
B_o=B_m \sim 3\times 10^{18} (\xi /30)~\G~, \eqno(4.6)
$$
where, in terms of the first of equations (4.1), $\xi\sim 30$. Now, however,
$\eta_r$ is no longer defined as $(1-\xi)\eta$
but rather just as $\eta_r=E_{ro}/M_oc^2$. The dynamic considerations are
similar to those just described, but the dynamics is entirely described
by using $\eta_B$ instead of $\eta$ everywhere in \S 3, and increasing
$\Gamma_m$ in eq.(3.14) by a factor 10 (since $E_{51}\sim 10^3$).
\bsk
{\it 4.2~~ Standard Shock Deceleration Model}
\msk

The magnetic fields are carried on in the comoving frame of whatever
polluting baryons $M_o$ were present originally in the fireball, and
continue expanding with the latter and the corresponding electrons plus
surviving pairs until the fireball matter is decelerated by the external
medium (Rees and \Mesz, 1992). For a fireball which has saturated its bulk
Lorentz factor to a value $\Gamma_s=\hbox{min}[\eta,\Gamma_m]$ the deceleration
occurs when the mass of external matter swept up by the fireball equals
$\Gamma_s^{-1}M_o$; the swept-up external matter is shock heated to a comoving
average thermal Lorentz factor $\gamma_s\sim \Gamma_s\sim \eta >>1$, while
a reverse shock starts to propagate into the adiabatically cooled fireball
material, eventually reheating it to a marginally relativistic average thermal
Lorentz factor $\gamma_r \sim \Gamma_r \siml 2$. If the forward and reverse
shocks
are able to create
a field which approaches equipartition with the post-shock thermal energies
of the particles (similarly to what appears to occur in supernova remnant
shocks), both the forward and reverse shock regions can radiate away the entire
thermalized bulk kinetic energy of the shock, i.e. the entire total initial
disposable energy $E_o$ of the event. However, because of the dependence of
the radiative efficiency on the magnetic field strength, it is worthwhile
to consider departures from shock equipartition, as well as the effect of
frozen-in primordial fields discussed in the previous subsection, whose initial
strength $B_o$ may have been amplified by shear or turbulent dynamo effects
during the cataclysmic event of the initial energy release, parametrized
through $\xi$ defined in the first of equations (4.1).

In order to estimate the radiative efficiency at the deceleration shock, we
use the parameters of our ``standard" shock model, e.g. \Mesz~ and Rees,
(1993). For typical parameters, the shock deceleration occurs after the
fireball has saturated (and after it has become optically thin, producing
a brief and weak precursor burst). The deceleration radius $r_d$ at which the
fireball gas starts to feel the inertia of an external medium of uniform
density
$n_{ext}=n_1\cmcui$ is $r_d/r_o= (\eta_{ext}/\eta^2)^{1/3}$, where $\eta_{ext}=
(E_o/V_o n_{ext} m_p c^2)$, or $r_d=(3 E_o/4\pi n_{ext}m_pc^2 \eta^2)^{1/3}$,
which is
$$
r_d \sim 10^{16}n_1^{1/3} E_{51}^{1/3} \eta_3^{-2/3}\cm~.\eqno(4.7)
$$
(For maximal magnetic dominance, $E_{51}\sim 10^3~,~ \eta\equiv \eta_B~,~
\xi\sim 30$, but for the purposes of this example we consider the more
conservative case of $E_{51}~\sim 1~,~ \xi\siml 1~,~n_1\sim 1$).
The comoving width of the forward and reverse shocked shells is $\eta$ times
larger than their laboratory width $\sim r_d/\eta^2$, or $\Delta R \sim
r_d/\eta
\sim
10^{13}\eta_3^{-5/3}\cm$. The comoving expansion time is consequently
$$
 t_{ex}=\Delta t_c=\Delta R/c \sim 10^3\eta_3^{-5/3}~\s~,\eqno(4.8)
$$
while the laboratory expansion time (the observed GRB burst time, if the shock
is radiatively efficient) is $\Delta t=r_d/\eta^2\sim 1~\eta_3^{-8/3}\s$. The
total
number of baryons (protons) involved in the fireball is $N_{p,o}=E_o/\eta
m_pc^2
\sim 10^{54} \eta^{-1}$, and the comoving density in the preshocked fireball
gas
(for $r_d >r_b$ in eqs.(3.8) with (3.4)) is
$$
n=n_d=N_{p,o}\eta/4\pi\theta^2 r_d^3\sim 10^6\eta_3^2\cmcui~.\eqno(4.9)
$$
If the reverse shock, whose bulk Lorentz factor achieves at most a value
$\Gamma_r\sim 2$, produces a shock equipartition field, this is
$B_{d,e}=(8\pi nm_pc^2\Gamma_r 4)^{1/2}\sim 10^2\eta_3\G$, while if the
magnetic field is mainly the original, possibly amplified and adiabatically
expanded fireball magnetic field (4.4) or (4.6), this is
$$
B_d=4 B_o\eta^{2/3}(r_d/r_o)^2=4\times 10^{-1}\xi\eta_3^2 \G~,\eqno(4.10)
$$
where in both cases we have included a factor 4 to take into account
the shock compression factor. Aside from the different $\eta$ dependence,
we may consider both the shock-equipartition and frozen-in fields to be given
by (4.10), where the shock-equipartition field case has $\xi\sim 250
\eta_3^{-1}$ while the frozen-in field has $\xi\siml 1$ (or $\xi\siml 30$ in
the extreme magnetic dominance case).
\bsk
{\it 4.3~~ Radiative Efficiency and Photon Energies}
\msk

The efficiency of the blast wave moving ahead of the contact discontinuity
can only be affected by the field $B_{d,e}$ developed in the shock, as
discussed in our previous papers. Here we focus on the radiative efficiency of
the reverse-shocked fireball material, which can be affected by the original,
possibly amplified, fireball field $B_d$ of eq.(4.10). The ``fireball" at the
stage just before the reverse shock moves into it contains only a small
fraction
of the cooled original pairs and the original polluting baryons and electrons.
With a field of order (4.10), relativistic electrons of Lorentz factor $\gamma
\sim 10^6$ produced by the reverse shock are sufficient to ensure a synchrotron
cooling time comparable to the (comoving) expansion time (4.6), i.e. near unit
radiative efficiency, and photons of characteristic energy $\sim 1 \eta_3 \MeV$
in
the laboratory frame. If shock acceleration is considered, limited by
synchrotron
losses, the maximum possible $\gamma\siml 2\times 10^7 B^{-1/2}$ and the
maximum
energy of the synchrotron photons is $\sim 1 \eta_3 \GeV$ in the laboratory
frame,
independent of the field strength.

Synchrotron losses, however, will almost certainly be surpassed in importance
by the inverse Compton (IC) losses of the electrons, in the radiation field of
the synchrotron photons. Without diffusive Fermi-type  acceleration , the
electrons passing through the shock may energetically attain
a maximum average Lorentz factor
$$
\gamma \sim (m_p/m_e)\Gamma_r\sim 4\times 10^3 \zeta~.\eqno(4.11)
$$
With this average Lorentz factor, the ratio of synchrotron cooling to expansion
time $t_{sy}/t_{ex}\sim 5\times 10^8\gamma^{-1}B^{-2}\sim 2.5\times 10^3
\zeta^{-1}\xi^{-2}\eta_3^{-7/3}\gg 1$, i.e. the synchrotron efficiency is low.
The energy density in synchrotron photons is $u_{sy}\sim n P_{sy} \Delta
R/c\sim
2\times 10^2 \zeta^2\eta_3^{1/3} u_B$, where $u_B=B^2/8\pi$ is the magnetic
energy density. The ratio of inverse Compton cooling time to expansion time
is therefore
$$
{t_{IC}\over t_{ex}}={u_B\over u_{sy}} \sim
10^1\eta^{-3}\xi^{-2}\eta_3^{-8/3}~,
                                                                   \eqno(4.12)
$$
giving a first-order inverse Compton efficiency of 10\% even for this modest
average Lorentz factor $\gamma \sim 4\times 10^3\zeta$ and $\xi\sim 1$. The
synchrotron photons (IR in the comoving frame, optical in the lab frame for
this
$\gamma$ and $\eta_3\sim 1$) are boosted up by the single inverse Compton
scattering to MeV and GeV energies in the comoving and lab frames.

The radiative efficiency will be larger in the presence of a diffusive
(or similar) shock acceleration mechanism that produces an electron power-law
distribution. Considering the electron shock acceleration to be limited by
first-order inverse Compton losses (using the Thomson limit in the electron
rest frame and balancing $t_{IC}$ with the acceleration time $t_a\sim \gamma
/\omega_B$ where $\omega_B=eB/m_e c$) one gets a maximum $\gamma$ significantly
higher than (4.11) by a factor $\zeta_m\sim 4.5\times 10^1\xi^{-1/4}
\eta_3^{-7/12}$. The synchrotron photon energy in the electron rest frame
becomes of order 1 MeV at a slightly lower energy, $\zeta_m\sim 2\times 10^1
\xi^{-1/3}\eta_3^{-2/3}$. Since the transition to the Klein-Nishina cross
section
at this energy will effectively limit the maximum Lorentz factor to this
value, we will use this second value of $\zeta_m\sim 20$, i.e. $\gamma_m\sim
10^5$. With this, the synchrotron and first-order IC photons have a
characteristic maximum energy of
$$
E_{sy}\sim 10 \xi^{1/3}\eta_3^{5/3} \keV~~,~~
E_{IC}\sim 10^{14}\xi^{-1/3}\eta_3^{1/3}\eV~,\eqno(4.13)
$$
in the lab frame, and a radiation efficiency
$$
\epsilon_{sy}\sim 10^{-2} \xi^{5/3}\eta_3^{5/3}~~,~~
\epsilon_{IC}\sim \hbox{min}[1~,~10^3\xi\eta_3^{2/3}]~.\eqno(4.14)
$$

Another effect which could increase the efficiency is the higher order
IC scattering (e.g. Rees, 1967). The Thomson optical depth of the fireball
shell material is
$$
\tau_T\sim n\sigma_T f\Delta R\sim 0.6\times 10^{-5} f
\eta_3^{1/3}~,\eqno(4.15)
$$
where $f\leq 1$ is the fraction of the fireball material that has been heated
by the shock. When $\tau_T$ is larger than $\gamma^{-2}$ (as long as the
scattering is not Klein-Nishina dominated) the conditions are fulfilled
for both first order and higher order inverse Compton scattering to dominate
over synchrotron losses. This is because the synchrotron,
first-order IC, second-order IC, etc. photon energy densities are
$$
\eqalignno{
u_{sy}\sim n P_{sy}f\Delta R/c  \sim n\sigma_T \gamma^2 u_B f\Delta R
&\sim \tau_T\gamma^2 u_B~,& (4.16a)\cr
u_{IC}\sim n P_{IC}f\Delta R/c  \sim n\sigma_T\gamma^2 u_{sy} f\Delta R
& \sim \tau_T^2\gamma^4 u_B~,& (4.16b) \cr
u_{IC'}\sim n P_{IC'}f\Delta R/c  \sim n\sigma_T\gamma^2 u_{IC}f\Delta R
&\sim \tau_T^3\gamma^6 u_B~,...&(4.16c) \cr}
$$
and for $\tau_T\simg \gamma^{-2}$ the ratio of the successive energy densities
in eq(4.16) are larger by increasing factors of $\tau_T\gamma^2 \geq 1$.
Even for $\gamma\sim 10^3$ the IC losses exceed synchrotron losses by a factor
10, and the second order IC' losses would exceed first order IC losses by
another factor of 10, were it not for the fact that for second order IC
and the {\it maximal} $\gamma$ the photons in the electron rest-frame are
now subject to the Klein-Nishina decrease in the cross section.
(For $\gamma\sim 10^5$ the respective factors are $10^5,~10^{10},$, etc.,
and the importance of the IC losses make themselves felt already when the shock
has heated a small fraction $f\simg 10^{-5}$ of the fireball material).
Notice that for both our simplest model (using the
average Lorentz factor $\gamma\sim 10^3$) and for the non-thermal (diffusive
acceleration) model with $\gamma_m\sim 10^5$, the first order IC is in the
Thomson limit and exceeds synchrotron losses, but the Klein-Nishina losses
diminish the importance (at least as far as the efficiency is concerned)
of the second and higher order IC scattering, although they will affect the
spectrum. In the power-law spectrum model, the first-order IC efficiency
is unity for $\xi\eta_3^{2/3} >10^{-3}$, i.e. for initial fields $B_o\simg
10^{14} \eta_3{-2/3}\G$ in the fireball, or for fields developed by turbulent
instabilities in the reverse shock which are a factor $10^{-5}\eta_3^{-2/3}$
below equipartition with the thermal (reverse) post-shock particles, a
fairly undemanding assumption
\bsk
\ctl{\bf 5.~~ Discussion}
\ssk

We have presented both analytical and numerical calculations of the evolution
of a freely expanding ultrarelativistic gas produced by an impulsive energy
release. The expansion and cooling from an initially optically thick and
extremely hot fireball is followed through the stage where the rest-frame
energy
density becomes comparable to the rest mass density, where the expansion bulk
Lorentz factor, which until then grew proportionally to the expansion factor,
saturates to a value which is equal to the smallest of the ratio of the initial
thermal energy to rest mass energy in the fireball or the bulk Lorentz factor
at which Compton drag becomes negligible. The laboratory and comoving frame
geometry as well as the radiation and gasdynamic variables were investigated
for a large range of values of $\eta=E_o/M_oc^2$, and we derived analytic
scaling laws showing the dependence of the variables as a function of the
various initial parameters, including an approximate angular scaling for the
case of jet-like expansion within a limited range of solid angles.

We have also devoted particular attention to the likely magnetic field
content of fireballs.  An initial magnetic field, possibly amplified by the
violent mass motions in the initial impulsive
event, may contribute a dynamically-significant  fraction of the total fireball
energy; there is also the  possibility that a field develops in the shock
heated
fireball after the ram pressure of an external medium decelerates the
expansion.
    When the fireball runs into an external medium and is decelerated, the
efficiency with which the re-randomized energy is radiated depends on the
magnetic field strength, and on  how the electrons are accelerated by the
resultant shocks. (The radiation processes during the deceleration phase which
give rise to the typical burst profiles are, fortunately, rather less sensitive
to  precisely how the fireball was originally formed --  the  formation
mechanism may have resembled the impulsive model discussed in sections 2 and 3,
or could alternatively have been spread over as long as a second, as in the
scenarios of Usov (1992) or Narayan et al (1992) ). We find, for a range of
assumptions,  that inverse synchro-Compton cooling should be extremely
efficient in radiating away the kinetic energy of the cooled fireball in a
time scale short compared to the expansion time, which gives the right order
of magnitude duration and total energy for a gamma ray burst source, and
produces a nonthermal spectrum with the bulk of the energy in the the MeV to
GeV range.
\bsk
This research has been partially supported through NASA NAGW-1522.
\msk
\ctl{\bf References}
\ref Benz, W., 1984, Astron. Astrophys., 139, 378
\ref Blandford, R.D. and McKee, C.F., 1976, Phys. Fluids, 19, 1130
\ref Cavallo, G. and Rees, M.J., 1978, M.N.R.A.S., 183, 359
\ref Courant, R., Friedrichs, K. O., and Lewy, H., 1928, Math. Ann., 100, 32
\ref Fishman, G., 1992, in {\it Proc. Saint Louis Compton GRO Symp.}, in press
\ref Goodman, J., 1986, Ap.J.(Lett.), 1986, 308, L47
\ref Imamura, J.N. and Epstein, R.I., 1987, Ap.J., 313, 711
\ref Meegan, L.A., Fishman, G.J., Wilson, R.B., Paciesas, W.S., Brock, M.N.
 Horack, J.M., Pendleton, G.N. and Kouveliotou, C., 1992, Nature, 335, 143.
\ref \Mesz, P. and Rees, M.J., 1992a, Ap.J., 397, 570.
\ref \Mesz, P. and Rees, M.J., 1992b, M.N.R.A.S., 257, 29P.
\ref \Mesz, P. and Rees, M.J., 1993, Ap.J., in press
\ref Narayan, R., Paczynski, B. and Piran, T., 1992, Ap.J.(Letters), 395, L83
\ref Paczy\'nski, B., 1986, Ap.J.(Lett.), 308, L43
\ref Paczy\'nski, B., 1990, Ap.J., 363, 218.
\ref Paczy\'nski, B., 1992, in {\it Gamma-ray Bursts}, eds. Paciesas, W. and
 Fishman, G. (A.I.P. Conf.Proc. 265, New York), p. 144
\ref Rees, M.J., 1967, M.N.R.A.S., 137, 429
\ref Rees, M.J. and \Mesz, P., 1992, M.N.R.A.S., 258, 41P.
\ref Shemi, A. and Piran, T., 1990, Ap.J.(Lett.), 365, L55
\ref Thompson, C. and Duncan, R.C., 1993, Ap.J., in press
\ref Usov, V.V., 1992, Nature, 357, 472
\ref Vitello, P. and Salvati, M., 1976, Phys. Fluids, 19, 1523.
\ref Woosley, S., 1993, Ap.J., in press
\ref Zdziarski, A.A., 1982, in {\it Accreting Neutron Stars}, eds. W. Brinkmann
 and J. Truemper (MPE Rep. 177, Max Planck Inst., Garching), p. 246.
\newpage
\ctl{\bf Figure Captions}
\bsk
\ref {\it Fig. 1} : Mass distribution in the laboratory frame for a fireball of
$\eta=10^4$ at the times when the lab expansion factor $r(\rho_{max})/r_o$ (
proportional to the lab time) is equal to $0.2, 0.4, 0.6, 0.8, 1.0$
times $\eta$, from top to bottom.
\ref {\it Fig. 2} : Distribution of the bulk Lorentz factor $\Gamma$ over the
various mass shells (i.e. against the Lagrangian mass coordinate), for
$\eta=10^4$,
at the times when the lab expansion factor ( which is proportional to the lab
time)
is $r(\rho_{max})/r_o=\eta^{1/2},\eta, \eta^{3/2}, \eta^2, \eta^{5/2},\eta^3$,
from
bottom to top.
\ref {\it Fig. 3} : The average bulk Lorentz factor $\Gamma$ as a function of
the expansion factor $r(\rho_{max})/r_o$, for $\eta=10^4$.
\ref {\it Fig. 4} : a) The lab frame mass density against the Lagrangian mass
coordinate, at the times (from top to bottom) when the expansion factor is
$r(\rho_{max})/r_o=\eta^{1/2},\eta, \eta^{3/2}, \eta^2, \eta^{5/2},\eta^3$.
\lbr
b) The comoving frame mass density against the Lagrangian mass coordinate, for
the same instants as a)
\ref {\it Fig. 5} : a) The lab frame width of the shell where most (80\%) of
the
mass is located, as a function of the expansion factor $r(\rho_{max})/r_o$, for
$\eta=10^4$. \lbr
b) The comoving frame width of the shell containing most (80\%) of the matter,
as a function of the expansion factor $r(\rho_{max})/r_o$, for $\eta=10^4$.
\end